# Pressure-induced structural phase transition and suppression of Jahn-Teller distortion in the quadruple perovskite structure

V. S. Bhadram[1*], B. Joseph[2], D. Delmonte[3], E. Gilioli[3], B. Baptiste[1], Y. Le Godec[1], R. P. S. M. Lobo[4,5], and A. Gauzzi[1*]

[1]*IMPMC, Sorbonne Université and CNRS, 4, place Jussieu, 75005 Paris, France*
[2]*Elettra Sincrotrone Trieste, S.S. 14, Km 163.5 in Area Science Park, Basovizza, Trieste 34012, Italy*
[3]*Istituto dei Materiali per Elettronica e Magnetismo-CNR, Area delle Scienze, 43100 Parma, Italy*
[4]*LPEM, ESPCI Paris, PSL University, CNRS, F-75005 Paris, France*
[5]*Sorbonne Université, CNRS, LPEM, F-75005 Paris, France*

By means of *in situ* synchrotron X-ray diffraction and Raman spectroscopy under hydrostatic pressure, we investigate the structural stability of the quadruple perovskite $LaMn_7O_{12}$. At 34 GPa, the data unveil a first-order structural phase transition from monoclinic $I2/m$ to cubic $Im\bar{3}$ symmetry characterized by a pronounced contraction of the unit cell and by a significant modifications in the Raman phonon modes. The phase transition is also marked by the suppression of Jahn-Teller distortion which is present in the ambient monoclinic phase. In addition, above 20 GPa pressure, a sudden and simultaneous broadening is observed in several Raman modes which suggests the onset of a sizable electron-phonon interaction and incipient charge mobility. Considering that $LaMn_7O_{12}$ is paramagnetic insulator at ambient, and Jahn-Teller distortion is frozen in the high-pressure $Im\bar{3}$ phase, we argue that this phase could be a potential candidate to host a purely electronic insulator-metal transition with no participation of the lattice.

## I. INTRODUCTION

Perovskite-like compounds are known to exhibit a variety of structural distortions as compared to the ideal cubic $Pm\bar{3}m$ symmetry of simple $ABO_3$ perovskites. These distortions, classified by Glazer[1], are stabilized by a complex interplay between elastic strain and charge, spin and orbital degrees of freedom, as discussed in the pioneering studies by Wohlan and Koehler[2] and by Goodenough[3] on doped mixed-valence manganites $(La,Ca)MnO_3$. Among all possible 23 distortion patterns classified by Glazer, the quadruple perovskite (QP) structure, first reported by Marezio *et al.* [4] and described by the chemical formula $(AA'_3)B_4O_{12}$, has attracted interest as it offers the opportunity of studying the rich physics of mixed-valence systems without chemical disorder and electronic inhomogeneities inherent to chemically substituted systems. The reason is a 1:3 ordering of two inequivalent A and A' sites generated by a very large tilt ($\psi \sim 137°$[5]) of the $BO_6$ octahedra around the [111] diagonal axis, denoted $a^+ a^+ a^+$ in Glazer's notation. This tilt is driven by a Jahn-Teller (JT) distortion of the A' site, where typically A'=$Cu^{2+}$ as in $(CaCu_3)Ti_4O_{12}$ [6] or $Mn^{3+}$ as in $(AMn_3)Mn_4O_{12}$[5,7]. Interestingly, this distortion turns the pristine dodecahedral coordination of the A' site into a rare square-planar coordination that prevents the formation of oxygen vacancies [8], a further favorable characteristic of QPs not found in simple perovskites. The above unique features may explain the fact that QP manganites $(AMn_3)Mn_4O_{12}$ exhibit almost full charge, spin and orbital orderings with sharp phase transitions and no coexistence of the disordered phase[7]. These manganites have attracted further interest for the record high values of electric polarizations induced by magnetic order reported for A=Ca, La [9] [10].

Here, we focus on a further feature of the $a^+ a^+ a^+$ tilt of the QP structure, namely the record high atomic packing among all possible tilt patterns of the pristine cubic $Pm\bar{3}m$ structure[11]. For the above tilt angle $\psi \sim 137°$, the density gain is as high as ~20% [8]. Such a maximization of packing prevents further pressure-induced lattice distortions, thus freezing the lattice degrees of freedom – a favorable situation for realizing a purely electronic pressure-induced insulator-metal transition with no participation of the lattice, as originally proposed by Mott [12]. A first high-pressure study on $(NaMn_3)Mn_4O_{12}$ corroborates the conjecture of the stability of the QP cubic $Im\bar{3}$ phase up to 20 GPa [13].

In order to investigate this possibility, in the present work we study the high-pressure behavior of the QP structure in $(LaMn_3)Mn_4O_{12}$ (LMO) using synchrotron powder X-ray diffraction (SXRD) and Raman spectroscopy. LMO is as a model system for our purpose owing to the following simple characteristics: 1) a paramagnetic and insulating ground state at ambient conditions; 2) single-valent $Mn^{3+}$ properties of the octahedral B site with no charge orderings [8]; 3) the high-symmetry cubic $Im\bar{3}$ structure with regular octahedra stable at high temperature undergoes a monoclinic $I2/m$ distortion at 653 K [14,15], concomitant to a JT distortion of the $MnO_6$ octahedra, which leads to an ordering of the $e_g$ orbitals of the $Mn^{3+}$ ions.

## II. EXPERIMENTAL METHODS

For the present experiment, high-purity LMO powder samples were synthesized using a high-pressure technique, as described elsewhere[15]. All the high-pressure SXRD and Raman experiments were performed at room temperature using diamond-anvil cells with rhenium (Re) gaskets and helium as pressure transmitting medium (PTM). This is crucial to minimize the non-hydrostatic stresses that may often modify the phase sequences in perovskites[16]. For SXRD measurements, a cluster of copper grains and a ruby crystal were loaded along with the sample. Pressure was estimated independently using ruby fluorescence and the equation of state of copper. SXRD measurements were carried out at the Xpress beamline of the Elettra synchrotron ($\lambda = 0.4957$ Å, beam size $\approx 30 \times 30$ μm$^2$) equipped with a MAR345 image plate detector[17]. The diffraction data were analyzed using the FULLPROF package[18]. Raman spectra were collected using a Horiba Jobin-Yvon HR-460 spectrometer equipped with a Andor charge coupled device (CCD) detector with a low-frequency cutoff of 120 cm$^{-1}$. A 514.5 nm Ar laser from Spectra Physics was used as an excitation source and an exposure time of 20 min was used for data acquisition. The Lorentzian function was used to fit the Raman modes.

## III. RESULTS AND DISCUSSION

### A. X-ray Diffraction

The SXRD pattern of LMO at ambient pressure is shown in Fig.1a. The pattern was easily indexed and refined to monoclinic cell with space group $I2/m$ and the obtained lattice parameters[19] are in close agreement with an earlier report[15]. There are weak impurity peaks present in the pattern (marked with an asterisk) which are unidentified but are probably arising from residual amounts (< 3%) of Mn$_3$O$_4$ and/or Mn$_2$O$_3$ known to be common impurity phases in LMO samples[8,14]. The inset of Fig.1a shows the characteristic Bragg peaks from $I2/m$ symmetry of the LMO phase. Due to the high compressibility of pressure medium (He), the Re gasket hole shrunk rapidly with pressure and the diffraction of gasket from the tail of the X-ray beam start to appear [19]. At pressures above 34 GPa, the splitting of the characteristic peaks with $I2/m$ symmetry disappeared (see inset of Fig.1a).

The pattern at 35.5 GPa was indexed to a cubic cell with $Im\bar{3}$ symmetry which is the parent structure for undoped QP manganites. LMO undergoes the same structural transition from $I2/m$ to $Im\bar{3}$ symmetry above 655 K at ambient pressure. [14] As per the Glazer scheme[1], this structure corresponds to $a^+ a^+ a^+$ octahedral tilt pattern. While the $I2/m$ is marked by the elongated Mn–O bonds in the neighboring MnO$_6$ octahedra of the monoclinic cell due to JT distortion (See Fig.1b), the $Im\bar{3}$ structure is marked by the identical Mn–O bond lengths[14]. This suggests the absence of JT distortion in the cubic structure due to symmetry constraints. The phase transition is reversible as seen from the diffraction pattern collected during the decompression run [19].

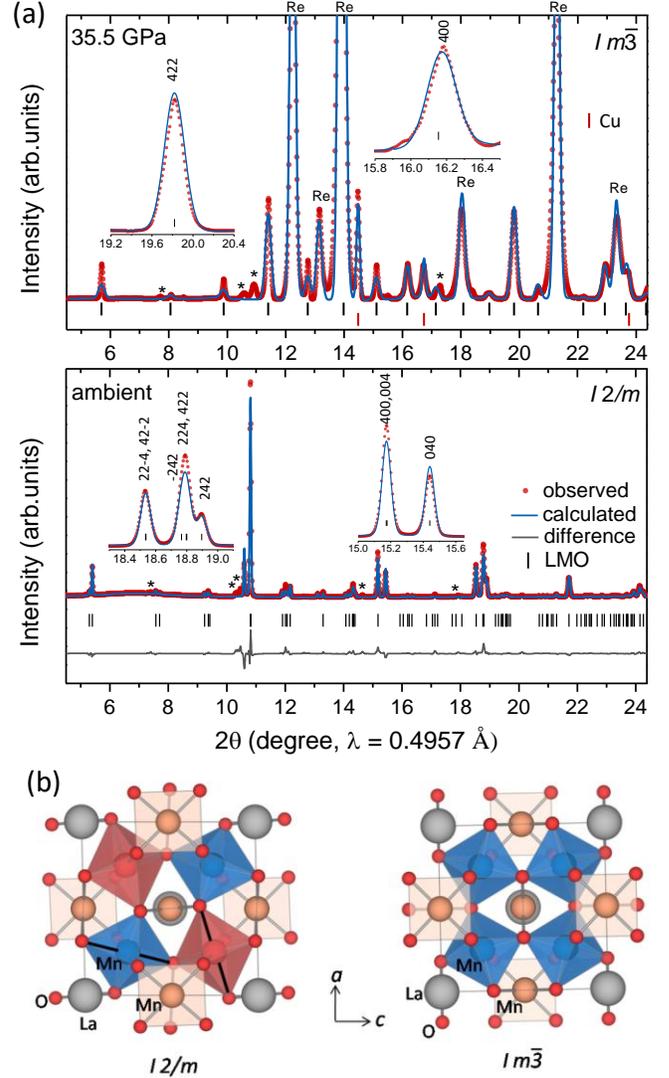

FIG. 1. (a) Refinement of the SXRD patterns of LMO at ambient (monoclinic I2/m phase) and at 35.5 GPa pressure (cubic Im$\bar{3}$ phase). The peaks originating from unidentified secondary phase(s) are denoted with asterisk (*). Insets show the Miller indices of characteristic regions. (b) Crystal structure of LMO in the monoclinic $I2/m$ (left) and cubic $Im\bar{3}$ (right) phases. Elongated Mn−O bonds due to the JT distortion in MnO$_6$ octahedra are marked with black lines.

It is intriguing that LMO undergoes the same structural transition with ascent of symmetry by increasing temperature or pressure. Indeed, the effect of pressure on lattice volume is rather analogous to that of low temperature, although the magnitude of volume change is quite different in the two cases. Thus, the structural phase transition from $I2/m$ to $Im\bar{3}$ may be of re-entrant type as it occurs during expansion and contraction of the lattice. Recently, Belik et al. [20] observed a low temperature re-entrant transition from room temperature rhombohedral $R\bar{3}$ to a parent cubic $Im\bar{3}$ phase in QP compound $BiCuMn_6O_{12}$. Similar to the present case of LMO, here the phase transition coincides with the collapse of the orbital degree of freedom of the Mn ions. On the other hand, it is also possible that the $I2/m$ to $Im\bar{3}$ phase transition at 655 K may have weak pressure dependence and thus occurring at room temperature at 34 GPa pressure. Further measurements at simultaneous high-pressure and high-temperature could be useful to prove this scenario.

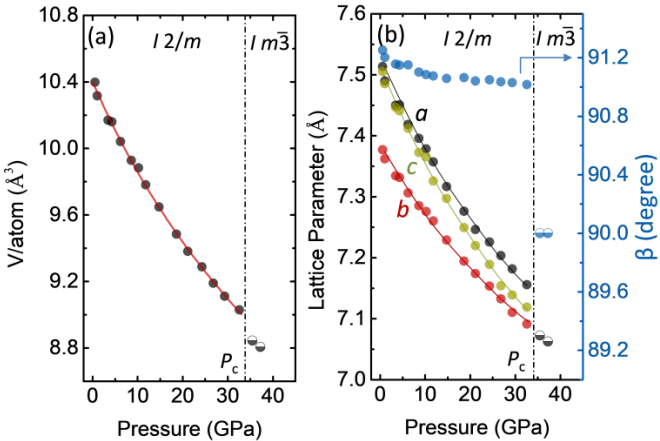

FIG. 2. Pressure dependence of (a) unit cell volume, (b) lattice parameters of LMO in monoclinic I 2/m and cubic Im $\bar{3}$ phases. Error bars in pressure, volume and lattice parameters are smaller than the symbol size. Solid lines are EoS fit to the experimental data (see text).

The lattice parameters as a function of pressure are shown in Fig. 2. All the lattice parameters decrease monotonously up to the transition pressure, $P_c$=34 GPa. A drop of 2.05 % in the lattice volume at $P_c$ indicates that the phase transition is of first order, contrary to the case of the transition at 655 K at ambient pressure where no discontinuity in the unit cell volume is observed[14]. The pressure-volume (V) data is fitted with 3$^{rd}$ order Birch-Murnaghan equation of state (EoS)

$$P(V) = \frac{3K_0}{2}\left[\left(\frac{V_0}{V}\right)^{7/3} - \left(\frac{V_0}{V}\right)^{5/3}\right]\left\{1 + \frac{3}{4}(K_0' - 4)\left[\left(\frac{V_0}{V}\right)^{2/3} - 1\right]\right\} \quad (1)$$

using EosFit software program [21]. The estimated isothermal bulk modulus, $K_0$ = 164.2(3) GPa (for first derivative, $K_0'$ = 4.54) is higher than that of simple perovskite $LaMnO_3$ ($K_0$ = 108 GPa)[22] possibly the result of compactness of the QP structure. Similar to the bulk modulus, the linear moduli ($M_0$) corresponding to cell axes are also obtained by using linear EoS available with EosFit program. They show a trend: $M_{0c}$ (452.5 GPa) < $M_{0a}$ (515.6 GPa) < $M_{0b}$ (669.2 GPa). This indicates that the compression is dominant along the longer $a$- and $c$-axes as compared to the short $b$-axis. This contrasts the high-temperature behavior where the expansion occurs predominantly along the $b$-axis[14].

Although the anisotropic behavior of the cell parameters under high-pressure is different from that at high-temperature, it leads to the same phase transition. This can be understood by looking at the behavior of octahedral Mn–O bond distances to the thermodynamic stimuli. According to a symmetry analysis of the monoclinic $I2/m$ distortion of the cubic $Im\bar{3}$ structure, the primary displacive modes transform according to the $\Gamma_4^+$ irreducible representation (Miller and Love notation). These atomic displacements lead to a cooperative JT distortion of the $MnO_6$ octahedra—the so called $Q2$ mode and the long-range B site $d_{3z^2-r^2}$ orbital order which sets in the $I2/m$ phase. Here, the octahedral elongation axis lies approximately within the monoclinic $ac$ plane and alternates from one site to the next as indicated in Fig. [1b]. At high-temperature, the unit cell expands anisotropically along the $b$-axis, eventually, leading to equal Mn–O bonds in the cubic $Im\bar{3}$ phase – suppression of JT distortion, and orbital order. Opposite to the high temperature behavior, the monoclinic cell contraction under pressure is dominant along the $a$- and $c$-axes. This means longer Mn–O bonds contract more rapidly than the shorter ones towards the suppression of the JT distortion and phase change to cubic $Im\bar{3}$. However, this requires an enormous pressure of 34 GPa whereas the same can be achieved at relatively moderate temperatures at ambient pressure. It is noteworthy that the stability of the JT distortion in a wide pressure range (0-34 GPa) in LMO is similar to its simple perovskite counterpart, $LaMnO_3$ [23] wherein JT energy associated with the low-symmetry distortion of the $MnO_6$ octahedron is $E_{JT} \approx 0.25$ eV/$Mn^{3+}$[24]. A simple model[24] shows that the maximum pressure required to suppress the JT distortion is directly proportional to $E_{JT}$ and that can be estimated from absorption spectroscopy or from the pressure dependence of $MnO_6$ octahedral volume which are currently not available for LMO.

## B. Raman Scattering

To gain further support to the structural phase transition observed in LMO and to know if there is any soft phonon mode possible driving the transition, Raman spectroscopy measurements were performed on a powder sample up to 44 GPa pressure as shown in Fig.3. To the best of our knowledge, the Raman response of LMO has not been reported in the literature. According to group theory, 18 Raman active modes (10 $A_g$ + 8 $B_g$) are expected for LMO in $I2/m$ symmetry. In our ambient pressure spectrum, we identify a total of 11 modes labeled $m_1$ to $m_{11}$ with increasing frequency, as shown in Fig.3a. We verified that none of these modes originate from the impurity phases[25]. Our observation of a lesser number of modes could either be due to accidental degeneracy; to an intrinsically weak intensity of modes; or both. Identifying the symmetry of each and every mode requires polarization dependent measurements on oriented crystal or lattice dynamical calculations which are beyond the scope of our present study. For the sake of completeness, we recall from the knowledge of Raman mode analysis of $LaMnO_3$[26, 27], the modes in the low-frequency region of the spectrum originate from the rotational motion of the $MnO_6$ octahedra whereas modes involving oxygen stretching motion would appear in the high-frequency region.

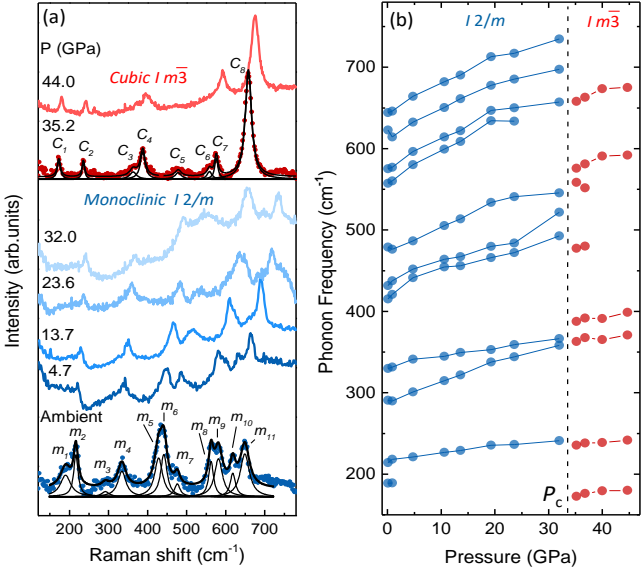

FIG. 3. (a) Unpolarized Raman spectra of LMO at different pressures. Deconvolution of the background corrected spectrum in $I2/m$ phase (at ambient) and $Im\bar{3}$ phase (35.2 GPa) are shown. (b) Pressure dependence of Raman phonon frequencies. Error bars are smaller than the symbol.

We now analyze the evolution of the Raman modes with pressure by excluding the low-frequency mode $m_1$ at 188 cm$^{-1}$,

no longer visible already in the 4.7 GPa spectrum. The main result is a striking change occurring in the spectrum between 32 and 35.2 GPa. Several modes disappear and some new modes appear indicating a structural phase transition which agrees well with the SXRD data. In the high-pressure cubic $Im\bar{3}$ phase, only 8 modes could be seen, labeled as $c_i$ ($i$ = 1 to 7). Observation of lesser number of modes is in accordance with the higher symmetry of the high-pressure phase as found in the SXRD analysis. The spectral features of the high-pressure phase qualitatively resemble those of the high-temperature cubic $Im\bar{3}$ phase of other QP manganites [28,29]. In this phase, only the oxygen ions participate in the 8 Raman active modes: $2A_g + 2E_g + 4F_g$. A strange broad background for the spectral region above 400 cm$^{-1}$ is observed at pressures leading up to the phase transition and it disappears in the $Im\bar{3}$ phase. As this spectral region corresponds to the octahedral Mn–O vibrations, the broad background could be originating from the fluctuations in JT distortion prior to its disappearance in the $Im\bar{3}$ phase. Nevertheless, the changes in the Raman spectra across the phase transition are reversible and the spectral features in the recovered sample match well with those of the starting sample [19].

The pressure dependence of mode frequencies is plotted in Fig [3b]. All the modes harden with pressure. While several modes disappear at the phase transition, a few new modes start to appear. The estimated Grüneisen parameters for individual modes ($\gamma_i = \frac{K_0}{\omega_0} \frac{d\omega_i}{dP}$) are listed in table I. None of the modes exhibit negative $\gamma_i$ suggesting the absence of any soft phonon mode or structural instability driving the phase transition. Although there are not enough points to estimate the $\gamma_i$ values in the $Im\bar{3}$ phase, Fig. 3(b) suggests that these values are similar in both phases.

**Table I:** Raman phonon frequencies and their Grüneisen parameter ($\gamma_i$) in the monoclinic $I2/m$ phase.

| Mode | $\omega_0$ (cm$^{-1}$) | $\gamma_i$ (Grüneisen Parameter) | Mode | $\omega_0$ (cm$^{-1}$) | $\gamma_i$ (Grüneisen Parameter) |
|---|---|---|---|---|---|
| $m_2$ | 214.12 | 1.144 | $m_7$ | 478.93 | 0.241 |
| $m_3$ | 290.76 | 1.073 | $m_8$ | 557.31 | 1.129 |
| $m_4$ | 329.87 | 1.108 | $m_9$ | 575.24 | 1.072 |
| $m_5$ | 415.35 | 2.453 | $m_{10}$ | 622.90 | 0.518 |
| $m_6$ | 431.53 | 1.926 | $m_{11}$ | 644.21 | 1.023 |

The pressure dependence of the mode linewidths is shown in Fig. 4. Remarkably, all modes, except modes $m_2$ and $m_9$,

display a steady decrease of linewidth with pressure followed by a sudden and simultaneous broadening above 20 GPa. In the absence of any structural changes around 20 GPa, such large line broadening in the Raman modes could be either due to intrinsic effects or the external factors such as deviatoric stresses in the sample chamber or on the sample. Even though helium is a good pressure medium, direct contact between sample and the gasket could develop anisotropic stress on the sample. However, any such stress on the sample could result in line broadening in all the modes. The observation that the line width of $m_2$ and $m_9$ modes is continuously decreasing even above 20 GPa clearly suggests that the observed line broadening in other modes is intrinsic. Moreover, we did not observe any anomalous changes in the mode frequencies around 20 GPa which further supports that the origin of line broadening is more of fundamental in nature rather than due to any external factors.

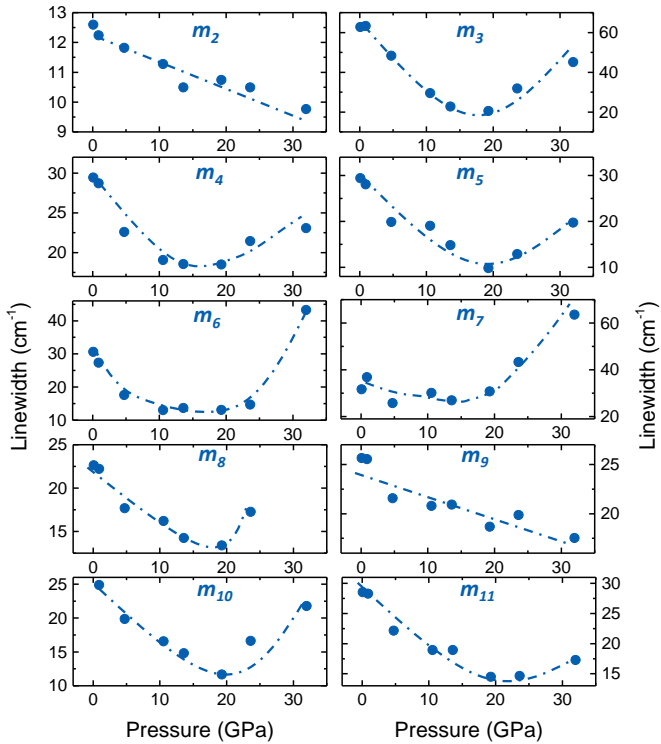

FIG. 4. Pressure dependence of Raman mode linewidths in the I2/m phase. Error bars are smaller than the size of the symbol. The dash-dotted line is guide to the eye.

While a quantitative explanation to understand the pressure dependence of linewidth of each mode would require *ab initio* calculations including anharmonic effects [30,31] and electron-phonon coupling [23], the fact that the above change of trend is common to almost all modes strongly suggests a marked change in the elastic and electronic properties of the system at 20 GPa. Namely, the line broadening observed above 20 GPa in the majority of the phonon modes indicates a sizable electron-phonon coupling, and thus a progressive increase of carrier density, suggestive of an incipient metallic phase. The question is whether this increase is sufficient to reach the critical carrier density required to stabilize a metallic phase and thus to drive the system across an insulator-metal transition. Although, we do not have any evidence to suggest such a transition occurring below 45 GPa, it would be an interesting scenario if it occurs at much higher pressures within $Im\overline{3}$ phase. Considering that LMO is a paramagnetic insulator at ambient conditions and that the JT distortion is frozen in the high-pressure $Im\overline{3}$ phase, the existence of an insulator-metal transition in the $Im\overline{3}$ phase would indicate a purely electronic insulator-metal transition with no participation of the lattice, as originally proposed by Mott[12]. To throw more light on such a scenario of incipient metallic phase, further *in situ* high-pressure transport property and *ab initio* studies would be needed.

## IV. CONCLUSION

We investigated the stability of the quadruple perovskite structure of LMO under high pressure. At 34 GPa we observed a first-order structural phase transition from monoclinic $I2/m$ to cubic $Im\overline{3}$ symmetry. The same ascent of symmetry is observed at ambient pressure upon increasing temperature. Our data show that the maximum compression of the unit cell is along the longer *a*- and *c*-axes, which means that pressure tends to equalize the Mn–O bonds in the MnO$_6$ octahedra, leading to the suppression of the JT distortion present in the $I2/m$ structure. The line broadening observed simultaneously in most Raman modes above 20 GPa suggests the appearance of sizable electron-phonon coupling and incipient charge mobility which may likely lead to a metallic phase at much higher pressures. Given that the high atomic packing of quadruple perovskite structure and the absence of JT distortion in the high-pressure $Im\overline{3}$ phase favors the freezing of the lattice degrees of freedom, any pressure induced metallic state in this phase would be a purely electronic insulator-metal transition or Mott transition. *In situ* transport property studies under high pressure and *ab initio* calculations would be needed to investigate this scenario further.

## Acknowledgements

The authors are grateful to M. Calandra and L. Paulatto for stimulating discussions, A. Polian and P. Parisiadis for assistance with the gas-loading apparatus and K. Beneut for assistance in the Raman measurements. They gratefully acknowledge the financial support by the French Government within the frame of the "Investissements d'Avenir" programme, Cluster of Excellence MATISSE, ANR-11-IDEX-0004-02.

*venkata.s.bhadram@gmail.com
*andrea.gauzzi@sorbonne-universite.fr